\documentclass[11pt,english]{article}

\makeatletter
\newcommand*{\rom}[1]{\expandafter\@slowromancap\romannumeral #1@}
\makeatother
\newcommand{\ket}[1]{\bigl| #1 \bigr\rangle}

\usepackage[latin9]{inputenc}
\usepackage{geometry}
\usepackage{verbatim}
\usepackage{prettyref}
\usepackage[shortlabels]{enumitem}
\usepackage{amsmath}
\usepackage{amssymb}
\usepackage{graphicx}

 \usepackage{slashed}
\usepackage{esvect}
\usepackage{dsfont}
\usepackage{tikz}

\usepackage{color}
\definecolor{darkgreen}{rgb}{0,0.5,0}
\definecolor{darkblue}{rgb}{0,0,0.6}
\definecolor{purple}{rgb}{0.4,.2,0.7}
\RequirePackage[colorlinks=true, urlcolor=black, menucolor=black, linkcolor=purple, citecolor=purple]{hyperref}
 	
\numberwithin{equation}{section}
\numberwithin{figure}{section}
\numberwithin{table}{section}

\def\CL{{\cal L}}
\def\tr{\,{\rm tr}\,}

\newcommand{\be}{\begin{equation}}
\newcommand{\ee}{\end{equation}}

\addtocounter{section}{0}

\title{\textbf{Apparently superluminal superfluids}}
\author{\href{mailto:ik2436@columbia.edu}{Ioanna Kourkoulou${}^{1}$}, \href{mailto:mjlandry@mit.edu}{Michael J. Landry$^{2}$},\\ \href{mailto:a.nicolis@columbia.edu}{Alberto Nicolis${}^{1}$}, and 
\href{mailto:k.parmentier@columbia.edu}{Klaas Parmentier${}^{1}$}\\\\\it \small ${}^{1}$\;Center for Theoretical Physics and Department of Physics,\\ \it \small Columbia University, New York, NY 10027, USA\\ \it\small ${}^{2}$\;Center for Theoretical Physics, Massachusetts Institute of Technology,\\ \it \small Cambridge, MA 02139, USA}
\date{}
\begin{document}
\fontseries{mx}\selectfont	
\maketitle
\begin{abstract}
We consider the superfluid phase of a specific renormalizable relativistic quantum field theory.
We prove that, within the regime of validity of perturbation theory and of the superfluid effective theory, there are consistent and regular vortex solutions where the superfluid's velocity field as traditionally defined smoothly interpolates between zero and arbitrarily large {\em superluminal} values. We show that this solution is free of instabilities and of superluminal excitations. We show that, in contrast, a generic vortex solution for an ordinary fluid does develop an instability if the velocity field becomes superluminal. All this questions the characterization of a superfluid velocity field as the actual velocity of ``something." 

\end{abstract}
\thispagestyle{empty}
\newpage
{\hypersetup{linkcolor=black}
\setcounter{tocdepth}{1}
\tableofcontents}
\addtocounter{page}{-1}
\vspace{0.5cm}
\section{Introduction }

Improving our understanding of motion  has arguably been one of the most important aspects of physics through the ages. We reached a peak with special and general relativity, but quantum mechanics immediately put a dent in the concept of a sharply defined trajectory for a point particle, by adding an irreducible uncertainty to position and momentum, and quantum field theory made the situation even worse, by making even the number of moving particles not particularly well defined in general. 

This is particularly manifest in relativistic hydrodynamics, where beyond the lowest order in a derivative expansion, there are so-called ``frame" ambiguities regarding how to define the local fluid velocity field \cite{Israel:1979wp}. For a non-relativistic fluid with only elastic collision processes, one can define the velocity field through kinetic theory, simply as the local statistical average of the invidual particles' velocities. But for a relativistic fluid such a definition is not particularly meaningful, and one instead focuses on local conserved currents, such as the stress-energy tensor and $U(1)$ charge current. Each of these  locally has a certain directionality and naturally defines a four-velocity field. It so happens, however, that different currents in general define different four-velocity fields. Which one is  the correct one? Does the question even make sense?
\newpage
The situation is somewhat ironic, in that relativistic quantum field theory has an absolutely sharp bound on motion -- microcausality: commutators and retarded two-point functions of local operators must vanish outside the lightcone, which we usually take as meaning that nothing can travel faster the light. Then, from this viewpoint, it seems that we can tell more easily how fast something {\it can}  {\it move} rather than how fast something  {\it is moving} or even {\it what} is moving.

With this paper, we want to exhibit yet another puzzle, regarding relativistic superfluids. We consider the theory of a complex scalar field with quartic interactions, invariant under a $U(1)$ symmetry, in a state of finite charge density. This, in four spacetime dimensions, is a renormalizable theory, and provides the simplest UV completion for the effective theory of a relativistic superfluid. We will show that at weak coupling and at small chemical potentials, there are stable vortex solutions for which the standard definition of the superfluid velocity field can  become arbitrarily superluminal. This happens close to the core of the vortex, but still well within the regime of validity of the superfluid effective theory. 

So, it appears that a superfluid can move faster than light. But all this is consistently derived within a renormalizable relativistic quantum field theory, where nothing can travel faster than light! In fact, we check that the {\em excitations} of our vortex solutions still obey microcausality: they are all subluminal even when the background appears to be superluminal. Our conclusion is that the standard definition of a superfluid velocity field 
might not correspond to the actual velocity of anything. Perhaps in the non-relativistic limit one can make physical sense of it, but for a relativistic superfluid it should not be taken literally as a velocity field. 

As a check, we consider a generic vortex configuration for an ordinary fluid. There, we show that as soon as the velocity field turns superluminal, the solution becomes unstable, with a UV-dominated instability rate, signaling that such a vortex is not a consistent solution within the fluid effective theory. We take this as an indication that, despite the frame ambiguities mentioned above, the four-velocity field of an ordinary relativistic fluid has a more physical status than that of a relativistic superfluid. 

\vspace{.5cm}

\noindent
{\it Notation and conventions:} We work in natural units ($\hbar = c = k_{ B} =1$) and  with the mostly-plus signature for the spacetime metric.

\newpage
\section{The fundamental and effective theories }



The simplest UV-completion for a relativistic superfluid's effective theory is given by a complex scalar field $\Phi$ with the $U(1)$ invariant action
\cite{Joyce:2022ydd}
\begin{equation}
    S = - \int |\partial\Phi|^2 + {\lambda}\big(|\Phi|^2-v^2 \big)^2 \; .
\end{equation}
We are interested in the $\lambda > 0$, $v^2 > 0$ case, which corresponds to having spontaneous symmetry breaking (SSB) already in the Poincar\'e invariant vacuum, as opposed to having it only for large enough chemical potentials \cite{Joyce:2022ydd}.
The reason will be clear in the next section.

Because of SSB, it is convenient to parametrize the scalar in polar field coordinates,
\be
\Phi (x)= \frac{\rho(x)}{\sqrt{2}}  e^{i \psi(x)} \; .
\ee

The angular mode $\psi$ is massless. On the other hand, the radial mode $\rho$ has mass of order $m^2 = 2 \lambda v^2$ and, at low energies  compared to $m$, one can integrate it out. At tree-level, this is equivalent to using its equation of motion, which to lowest order in derivatives reads
\begin{equation}
    \rho^2 \simeq 2 v^2 + \frac{1}{\lambda} X \; , \qquad X \equiv  -\partial_\mu \psi \partial^\mu \psi \; .
\end{equation}
Then, the effective low-energy Lagrangian for the Goldstone $\psi$ is \cite{Joyce:2022ydd, Babichev:2018twg}
\begin{equation}\label{eq:EFT1}
    \CL_{\rm eff}[\psi] \simeq \frac{1}{4\lambda} (2 m^2 X + X^2) \; , \qquad m^2 = 2 \lambda v^2 \; .
\end{equation}
This result is approximate in two senses:
\begin{enumerate}
\item In the effective field theory sense: it is the lowest order in the derivative expansion, for $\partial/m \ll 1$. Notice however that higher derivative corrections will not involve higher powers of $X = -(\partial \psi)^2$ without derivatives acting on them. This is because $\rho$ couples to $\psi$ only through the combination $X$. So, from the point of view of the derivative expansion, the effective theory above is correct to {\em all orders} in $X$, but to zeroth order in its derivatives (see a discussion in \cite{Joyce:2022ydd}).
\item In the small coupling sense: it is the lowest order in the perturbative expansion, for $\lambda \ll 1$.
More generally, to this order in derivatives, the $U(1)$-breaking pattern allows for
\begin{equation}\label{eq:EFT}
    \CL_{\rm eff}[\psi] =  P(X)\; ,
\end{equation}
with generic $P$.
In fact, at one-loop the result is of this form, with ${\cal O}(\lambda)$ corrections relative to \eqref{eq:EFT1} \cite{Joyce:2022ydd}. For our purposes, \eqref{eq:EFT1} will be enough.
\end{enumerate}

Now, in this theory a superfluid at equilibrium and at  rest in the lab frame can be thought of as a field configuration $ \psi(x)$ with constant time derivative, $ \psi(x) = \mu t$, where $\mu$ is the chemical potential. More in general, any field configuration $ \psi(x)$ with a nonzero $\partial_\mu  \psi$ and with mild gradients thereof (compared to $m$) can be thought of as a superfluid state, possibly featuring excitations or some nontrivial flow. 
 
More explicitly, one can consider the $U(1)$ current and the stress-energy tensor  associated with the general EFT \eqref{eq:EFT},
\be \label{JT}
J^\mu = 2 P'(X) \partial^\mu \psi \; ,  \qquad T_{\mu \nu} = 2 P'(X) \partial_\mu\psi \partial_\nu \psi +  P(X) \, \eta_{\mu\nu}  \; .
\ee
On the other hand, for an ordinary fluid, in the perfect fluid limit we would write
\be \label{JT2}
J^\mu = n \, u^\mu \; , \qquad T_{\mu \nu} = (\varrho + p) \, u_\mu u_\nu + p \,  \eta_{\mu\nu}  \; ,
\ee
where $n$, $\varrho$, and $p$ are the number density, energy density, and pressure, and $u^\mu$ is the fluid's four-velocity field.
By comparing \eqref{JT} with \eqref{JT2}, there seems to be no doubt that, if we want to associate some form of fluid motion with what we call a superfluid, we should identify the fluid four-velocity field with a suitably normalized version of $\partial_\mu \psi$,
\footnote{The overall sign is conventional. Our choice corresponds to $u^\mu$ being future-directed for positive $\partial_0 \psi$, which we identify with positive chemical potential and positive charge density.}
\be \label{eq:umu}
u_\mu = - \frac{\partial_\mu \psi}{\sqrt{|X|}} \; .
\ee  
It is also immediate to extract the values of $n$, $\varrho$, and $p$ associated with our superfluid. For what follows, it is actually more interesting to focus on the enthalpy density $\varrho+p$,  which for our specific UV completion reads
\be \label{rho + p}
\varrho+p = 2 P'(X) X \simeq \frac1\lambda (m^2 + X)X \; ,
\ee
where we restricted to the timelike $\partial_\mu \psi$ (positive $X$) case.

\newpage
\section{The superluminal vortex}\label{sec:vort}

We now come to our specific vortex solution. Within the effective theory, it is simply
\be \label{eq:vortex}
\bar \psi(x) = \mu t + \varphi \; ,
\ee
where $\mu$ is the asymptotic chemical potential at spatial infinity, which we take to be small,
\be
\mu \ll m \; ,
\ee
and $\varphi$ is azimuthal angle about an arbitrary axis (say, the $z$ axis.) It is immediate to verify that this $\bar \psi(x)$ obeys the EFT equations of motion for any $P(X)$,
\be
\partial_\mu \big( P(\bar X) \, \partial^\mu\bar \psi \big) = 0 \; .
\ee

This solution is singular at the $z$-axis, and one can ask if our UV completion resolves the singularity. As usual \cite{Weinberg:2012pjx}, this question can be phrased as an ODE for the radial mode with two boundary conditions, one at the $z$-axis and one at infinity, which always has a solution. In particular, close to the $z$-axis, at distances smaller than $r \sim m^{-1}$, the radial mode smoothly interpolates between its SSB minimum and zero, thus restoring the $U(1)$ symmetry at the vortex's center. 

The low-energy effective theory breaks down at such small distances, and this is something that one can infer just by looking at the solution. For example, on our solution we have
\be \label{bar X}
X(x) = \bar X(x) \equiv \mu^2 - \frac{1}{r^2} \; ,
\ee
where $r$ is the distance from the $z$-axis, and so
\be
\partial X (r \ll \mu^{-1}) \sim \frac{X}{r} \; ,  
\ee 
signaling that the effective theory breaks down at distances from the $z$-axis of order
\be
r_{\rm UV} \equiv m^{-1}
\ee
and shorter.

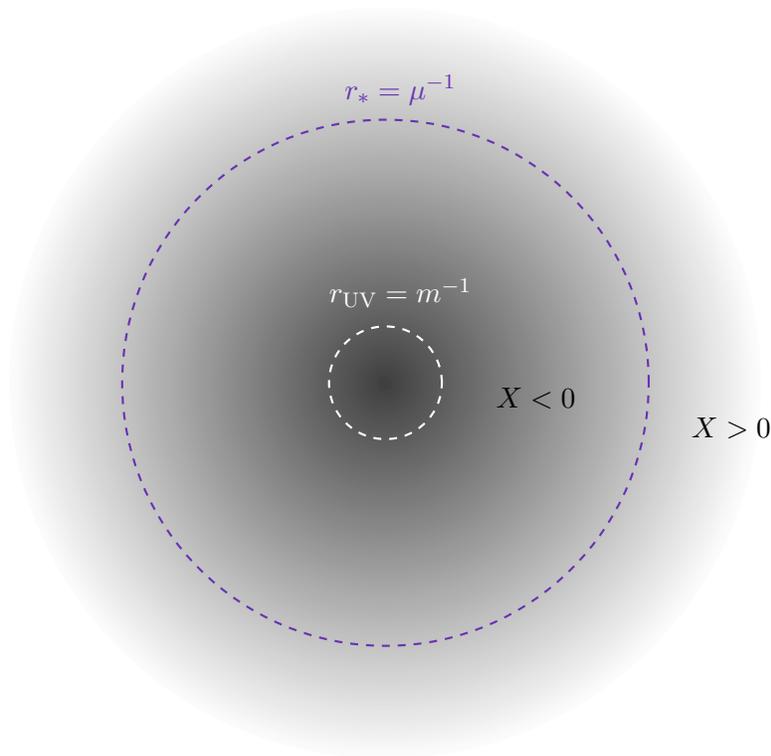
\begin{figure}[ht]
\centering
\begin{tikzpicture}[scale=1]
\shade[inner color=darkgray,outer color=white] (0,0) circle (5cm);
\node[purple] at (0.2,3.9) {$r_* = \mu^{-1}$};
\draw[purple,thick,dashed] (0,0) circle (3.5 cm);
\node[white] at (0.2,1.2) {$r_{\rm UV} = m^{-1}$};
\node[black] at (2.0,-0.2) {$X < 0$};
\node[black] at (4.6,-0.6) {$X > 0$};
\draw[white,thick,dashed] (0,0) circle (0.75 cm);;
\end{tikzpicture}
\caption{\small \it The vortex solution \eqref{eq:vortex} smoothly interpolates between timelike and spacelike superfluid velocities, as determined by the value of $X$ in \eqref{bar X}.  At $r_* = \mu^{-1}$, indicated by the purple dashed line, the background becomes lightlike. The EFT breaks down at scales smaller than $r_{\rm UV} = m^{-1}$, indicated by the white line. Fluctuations remain well-behaved in the intermediate region, where the background velocity formally becomes superluminal. }\label{fig:vortex}
\end{figure}

However, if we restrict to $\mu \ll m$, there is a much bigger critical scale,
\be
r_* \equiv \mu^{-1} \gg r_{\rm UV} \; ,
\ee 
where we can still trust the effective theory, but which however corresponds to an interesting transition: if we look at 
our $\bar X$ above (eq.~\eqref{bar X}), we discover immediately that right at $r=r_*$, $\bar X$ changes sign, going from positive to negative as one moves in from larger to smaller distances. Negative $X$ corresponds to spacelike $\partial_\mu \psi$ and thus to spacelike $u^\mu$, according to \eqref{eq:umu}. If $u^\mu$ is to be interpreted as the four-velocity of our superfluid, then our superfluid is moving superluminally for $r_{\rm UV} \ll r \ll r_*$. A visual representation of the vortex can be found in figure \ref{fig:vortex}.

Before addressing in detail the question of stability and causality of our solution, we want to emphasize a few points:
\begin{enumerate}
\item
If we abandon momentarily the superfluid interpretation, we realize that our solution is  nothing new. Indeed, setting $\mu$ to zero for a moment, our solution is nothing but the usual ``cosmic string" solution for theories with a spontaneously broken global $U(1)$ symmetry \cite{Weinberg:2012pjx}. Such configurations are well studied in the literature, and are known to be consistent QFT saddle points, which are, in particular, stable against small perturbations. Now, turning on  a tiny $\mu \ll m$ does not change in a substantial way the physics at energies much higher than $\mu$, or, equivalently, at distances $r$ from the core much smaller than $\mu^{-1}$, which is exactly the regime in which our superfluid becomes {\it very superluminal}. So, from this viewpoint, the superluminal regime corresponds to the {\it standard} physics of a cosmic string.

\item
Related to the previous point, we can now appreciate why we need to have SSB already at zero $\mu$: we want to start with the relativistic theory of a $U(1)$ Goldstone that makes sense if expanded about $X=0$. This  is because $X = 0$ is the  divide between subluminality and superluminality  in the superfluid interpretation, and we want to find a solution that interpolates between the two regimes.
Moreover, we want to work at $\mu^2 \ll m^2$, to that there is a wide range of scales, $r_{\rm UV} = m^{-1}\ll r \ll r_*=\mu^{-1}$ where our superfluid velocity is superluminal within the regime of validity of the effective theory.

\item
Our solution exists for a generic $P(X)$, and the resulting superluminality can be trusted for $\mu$ much smaller than the UV cutoff of that effective theory (the mass of the radial mode $m$, in our case). Because of this, our conclusions are robust against the inclusion of higher orders in perturbation theory. Whatever we prove using \eqref{eq:EFT1} will be qualitatively correct to all orders, and quantitatively it will receive corrections involving higher powers of $\lambda \ll 1$.

\item
Similarly, higher derivative corrections \eqref{eq:EFT1} will not change our conclusions qualitatively, but only quantitatively, by powers of $1/(r_{*} m) = \mu/m \ll 1$.

\item
Usually, within special relativity, one of the signs that it is impossible to accelerate a massive particle past the speed the light is that the amount of  energy needed to get closer and closer to the speed of light grows indefinitely. For a relativistic fluid substance, what measures local {\it inertia} -- what weighs kinetic energy -- is the enthalpy density $\varrho+ p$. In our case this is proportional to $X$ (see eq.~\eqref{rho + p}), and so the inertia of our superfluid  vanishes precisely when the superfluid velocity is crossing the speed of light, at $X = 0$, making it possible to cross that boundary. 

\item
While the $U(1)$ current $J^\mu$ becomes spacelike as soon as $u^\mu$ does, the four-momentum density $T^{0\mu}$ is timelike (or null) everywhere. It is not clear how much physical significance to assign to either of these two facts though. In general, a purely spatial $J^\mu$ can be set up by having an equal amount of positive and negative charges move opposite to each other at the same speed, and an arbitrary $T^{0\mu}$ can be made timelike or spacelike by adding a suitable cosmological constant, which, in the absence of gravity, has no physical consequence.
 
%
%
%

\end{enumerate}

 \newpage
\section{Stability and causality
}\label{sec:sup}
We now want to study the dynamics of small perturbations about our vortex solution.
Given the first two items at the end of the last section, we do not expect many surprises: we will be expanding the standard relativistic theory of a $U(1)$ Goldstone, eq.~\eqref{eq:EFT1}, about a weak background field. Since that theory is stable about a trivial background, and since the quartic interaction has the correct (positive) sign \cite{Adams:2006sv}, introducing a small background cannot affect the theory's stability or causality properties. 

However, given our superfluid interpretation, it is instructive to look at the quadratic action for the perturbations directly, and to see how these remain well-behaved even in the superluminal superfluid velocity regime. To this end, we will exploit the $r \gg m^{-1}$ hierarchy: at large distances from the core of the string, we can zoom in on a small patch of size $L$, still within the regime of validity of the effective theory, $L \gg m^{-1}$, but small enough so that the effects of the curvature of our vortex solution can be neglected, $L \ll r$. In this case, we can take for our Goldstone a linear background, 
\be \label{background Vmu}
\bar \psi (x)= V_\mu \, x^\mu +{\rm const} \; ,
\ee
and expand in small perturbations about it. This approach gives us the leading order dynamics for perturbations with wavelengths much smaller than $r$. The two cases $r > r_*$ and $r < r_*$ correspond, respectively, to timelike  $V_\mu$ and spacelike $V_\mu$.

Rewriting $\psi(x)$ as $\psi(x) = \bar \psi (x) + \pi(x)$, with $\bar \psi$ given by the background above, we get the quadratic action for the $\pi$ perturbation
\begin{equation}\label{eq:Lpiloc}
\mathcal{L}_\pi = \frac{1}{2\lambda} \big(( \,V^2 - m^2) \eta^{\mu\nu} + 2 V^\mu V^\nu \big) \partial_\mu \pi \partial_\nu \pi \equiv \frac12 Z^{\mu\nu}  \, \partial_\mu \pi \partial_\nu \pi
\; .
\end{equation}
A Lorentz-invariant condition for stability is \cite{Nicolis:2004qq, Kourkoulou:2022doz}
\be \label{weak stability}
Z^{00} > 0 \; , \qquad G^{ij} \succ 0 \; , \qquad G^{ij}=Z^{0i} Z^{0j}-Z^{00} Z^{ij} \; .
\ee
(The `$\, \succ\, $' symbol for a matrix means `positive definite'.)
In our case, since we trust the effective theory for our vortex solution only for $\partial \psi  = V \ll m$, those two conditions read
\be \label{simple stability}
1 + {\cal O}(V^2/m^2)> 0 \; , \qquad  \delta^{ij} + {\cal O}(V^2/m^2) \succ 0 \; ,
\ee
which are clearly obeyed. In fact, to quadratic order in $V_\mu$, the linearized equations of motion read
\be
\big(\eta^{\mu\nu} - 2 \,V^\mu V^\nu/m^2 \big) \, \partial_\mu \partial_\nu \pi \simeq 0 \; ,
\ee  
corresponding to the dispersion relation
\be
\omega^2 = \omega^2_{\vec k} \simeq |\vec k  |^2 - \frac{2}{m^2} \Big(V^0 |\vec k  | - \vec V \cdot \vec{ k}  \Big)^2  \; .
\ee
Clearly, within our $V \ll m$ approximation, the solutions for $\omega$ are real, thus showing that there are no exponentially growing modes. (For simplicity, here we are only displaying the positive frequency solutions. The negative frequency ones are related to these by a suitable sign change, $  \omega^-_{\vec k} = -  \omega_{ - \vec k} $.)

To check that excitations do indeed respect causality, we notice that the square of their direction-dependent phase velocity is
\footnote{It is easy to check that, to the order at which we are working, the square of the phase velocity agrees with the square of the group velocity, $ v^2_g ( {\hat k})= \big| \frac{\partial \omega}{\partial \vec k} \big|^2$. That is, the two notions of velocity differ only in their directions, but not in their absolute values.}
\be \label{phase velocity}
c_s^2 ({\hat k}) = \frac{\omega^2_{\vec k}}{| \vec k |^2}  \simeq 1- \frac{2}{m^2} \Big(V^0 - \vec V \cdot {\hat k}  \Big)^2 \; .
\ee  
Regardless of the value of $V^\mu$ and of the orientation of $\hat { k}$ relative to it, this can never exceed the speed of light, at best reaching it only for specific, $V^\mu$-dependent propagation directions 
\footnote{For spacelike and null $V^\mu$, the speed of light is attained in a direction $\hat k$ such that $\cos \theta \equiv \hat{ V} \cdot {\hat k} = V^0/|\vec V|$. For time-like $V^\mu$, the maximum speed is attained for $\hat k$ parallel or anti-paralled to $\vec V$ (depending on the sign of $V^0$), and is always subluminal, $c^2_{\max} = 1- \frac{2}{m^2} (|V^0| - | \vec V |  )^2 < 1$.}.
More carefully, one might want to check the theory's causal properties by inspecting the support of the position-space $\pi\pi$ retarded Green's function rather than simply the propagation speeds of free waves \cite{Dubovsky:2007ac}. Given the quadratic Lagrangian \eqref{eq:Lpiloc}, in Fourier space we have
\be
\tilde G_{\rm ret}(k) = \frac{i}{Z^{\mu\nu} k'_\mu k'_\nu} \; , \qquad k' {}^\mu \equiv (\omega + i \epsilon, \vec k) \; ,
\ee
and so in position space we find
\be
G_{\rm ret}(x) = \int \frac{d^4 k}{(2 \pi)^4} \, \frac{i}{Z^{\mu\nu} k'_\mu k'_\nu} \, e^{i k\cdot x} \; .
\ee
Treating $Z^{\mu\nu}$ as an effective inverse (constant) metric, defining a tetrad $e^\mu{}_\alpha$ through $Z^{\mu\nu} = e^\mu{}_\alpha e^\nu{}_\beta \, \eta^{\alpha\beta}$, and using standard GR manipulations, we  expect $G_{\rm ret}(x)$ to be
\be
G_{\rm ret}(x) = \frac{1}{\sqrt{-\det Z}} \,  G_0(y) \; ,
\ee
where $G_0(y)$ is the Lorentz-invariant expression, in Minkowski coordinates $y^\alpha = (e^{-1})^\alpha{}_\mu x^\mu$, for the retarded Green's function for a massless free scalar in a Poincar\'e invariant vacuum. Since this vanishes outside the light-cone, that is for $y^2>0$, our Green's function vanishes for $(Z^{-1})_{\mu\nu} x^\mu x^\nu < 0$, that is, outside a modified light-cone with a direction-dependent aperture given precisely by the phase velocity \eqref{phase velocity}. 
\newpage
\subsection*{Comments on Cherenkov emission}\label{sec:certime}
As discussed in \cite{Nicolis:2004qq, Dubovsky:2005xd, Kourkoulou:2022doz}, in certain situations one might want to impose a form of stability  stronger than \eqref{weak stability}, namely
\be \label{strong stability} 
Z^{00} > 0 \; , \qquad -Z^{ij} \succ 0 \; .
\ee
Such a condition is not Lorentz invariant, and there are physical situations in which it is violated in specific reference frames but not in others. Most notably, when a source is moving faster than the speed of sound relative to the medium it is moving in, in its rest frame the above condition is violated: the associated instability corresponds to the possibility of emitting Cherenkov phonons (or classical sound waves). This is a form of instability, but an instability that depends on the source. In particular, even though Cherenkov emission is peaked in the UV, $d \Gamma/ d \omega \sim \omega^2$ \cite{Nicolis:2017eqo}, it is cut off at a source-dependent frequency, the inverse sound-crossing time for the source. Moreover, it has a localized origin -- the source itself -- and so its effects are confined. As a result, even when present, Cherenkov emission is not a violent instability that destroys the system: for example, a fighter jet crossing the sound barrier does not set
the sky ablaze.

Still, in order to be absolutely sure that our vortex solution is viable, we want to see how it behaves as far as the condition \eqref{strong stability} is concerned. Since this condition is not Lorentz invariant, we have to choose a frame where to check it. We choose the `lab frame' where the vortex core is at rest; if we have stationary sources in our lab, that is the relevant frame where to check for Cherenkov emission. Since within the regime of validity of the effective theory all components of $V^\mu$ are small in that frame, $V \ll m$, the  condition \eqref{strong stability} still reduces to \eqref{simple stability}, which is trivially obeyed.
We conclude that our vortex solution is stable also against Cherenkov emission by sources that are stationary in the lab frame.

\newpage
\section{Instability for normal fluids}\label{sec:fluid}
We now consider a similar setup for regular fluids. In contrast to the superfluid case, we will see that gradient instabilities develop as soon as the fluid velocity field becomes superluminal.

From an EFT standpoint, a normal fluid, like a solid, is described by an $SO(3)$-triplet of scalars $\phi^I(x)$ for which spacetime translations $P^i$ are broken, but a combination of $P^i$ and internal shifts remains unbroken \cite{Nicolis:2015sra}. In the case of an isotropic solid, the action is $SO(3)$-invariant. 
Normal fluids constitute a special case thereof, with an action that is invariant under the larger group of 3-volume-preserving diffeomorphisms. This ensures that fluid elements can slide past each other without creating transverse stresses. The scalars $\phi^I(\vec x, t)$ can be thought of as the Lagrangian (i.e., comoving) coordinates corresponding to  Eulerian position $\vec x$ at time $t$. For a normal fluid, the low-energy EFT thus comes with the lowest order action
\begin{equation}\label{eq: fluid}
    S = \int \! d^4 x \, F\big(\det B\big) \; , \qquad B^{IJ}\equiv \partial^\mu \phi^I \partial_\mu \phi^J \; ,
\end{equation}
where $F$ is a function determined by the equation of state.
The normalized fluid 4-velocity satisfies $u^\mu \partial_\mu \phi^I = 0$ and is therefore determined to be
\begin{equation}
    u^\mu = \frac{1}{3!\sqrt{\det B }} \, \epsilon^{\mu \alpha \beta \gamma}\epsilon_{IJK} \, \partial_\alpha \phi^I \partial_\beta \phi^J \partial_\gamma \phi^K.
\end{equation}

We can now imagine that we have set up a vortex solution and that, as for the superfluid, there is a macroscopic critical distance from the axis below which the velocity field becomes superluminal. To study the local stability properties of such a solution, we can adopt the same approximation as for the superfluid case: at distances much bigger than the effective theory's UV cutoff $\ell$ (such as the mean free path for a weakly coupled gas), $r \gg \ell$, there is a range of scales $L$ where we can still use the effective theory, $L \gg \ell$ {\it and} we can neglect the effects of curvature of the vortex solution $L \ll r$. In that window of scales we can approximate the building blocks of the effective theory, $B^{IJ}$ and $u^\mu$, as constant, that is, we can take the background values for our fields $\phi^I$ as linear in $x^\mu$:
\be \label{background fluid}
\bar \phi^I (x)= a^I {}_\mu x^\mu \; ,
\ee
where  $a$ is a $3\times4$ matrix, and we neglected an irrelevant additive constant. We now want to perturb this solution and study its stability. To avoid clutter, we want to use the symmetries of the system. Both the action and the stability criterion \eqref{weak stability} (or its multi-field analog) are Lorentz invariant, so we can choose any frame that we find convenient. Moreover, the action is invariant under internal volume preserving diffs (acting on the $I$ index). We can simplify the expression \eqref{background fluid} while maintaining its linearity if we restrict to the linear subgroup of these transformations, that is, $SL(3, \mathbb{R})$. Each element of this subgroup can be thought of as a combination of a rotation and a shear transformation \cite{Nicolis:2022gzh}.

We start by performing a rotation in internal space ($I$ index) and a rotation in physical space ($\mu = i$) so as to diagonalize the $3 \times 3$ block $a^I {}_{i}$. Then, we can rescale the three $I=1,2,3$ axes with a volume-preserving shear so as to make $a^I {}_{i}$ proportional to the identity $\delta^I_i$. This is invariant under combined internal/spatial rotations, and so we can use those to align $a^I {}_0$ with the first direction in $I$ space. 
So far, we are left with 
\be
a^I {}_i = C \, \delta^I_i \; , \qquad a^I {}_0 = D \, \delta_1^{I}  \; .
\ee
Where $C$ and $D$ are two arbitrary numbers. 

We can now use boost invariance: if $C > D$, which corresponds to a subluminal $u^\mu$, we can boost to a frame where $D$ is zero. This keeps $a^I {}_i$ diagonal but makes it anisotropic. After performing an isotropizing shear in $I$-space, our fields are simply
\be
\phi^I (x) = \alpha \big(x^I + \pi^I(x) \big)  \qquad \qquad \mbox{(subluminal)}\; ,
\ee
where   $\alpha$ is an arbitrary constant and the $\pi^I$'s are small perturbations. As far the background is concerned, this is equivalent to going to the rest frame of the fluid, where $u^\mu = (1,\vec 0)$.
If on the other hand $D > C$, which corresponds to a superluminal $u^\mu$, we can boost to a frame where $a^I _1$ is zero, in which case, after a suitable shear transformation, our fields read
\be
\phi^1 (x) = \alpha  \big( t + \pi^1(x) \big) \; , \qquad 
\phi^{a=2,3} (x) = \alpha  \big(  x^a+ \pi^a(x) \big) \qquad \qquad \mbox{(superluminal)}\; .
\ee
For the background, this is equivalent to going to a frame where $u^\mu$ is purely spatial, $u^\mu = (0,1,0,0)$.

Plugging these expressions into the action, and expanding  in the $\pi$ fields (see e.g.~\cite{Dubovsky:2005xd}), we find the quadratric actions for the perturbations. In the subluminal $u^\mu$ case, we have the standard action for fluid perturbations in the fluid's rest frame \cite{Dubovsky:2005xd}
\begin{equation}
    \mathcal{L} = - \alpha^6 F' \Big[ \dot{\vec \pi} \,^2 - c_s^2 (\vec{\nabla} \cdot \vec{\pi})^2 \Big] 
     \qquad \qquad \mbox{(subluminal)}\; ,
\end{equation}
where the speed of sound is
\begin{equation}
    c_s^2 = \frac{d p}{d\varrho} = 1 + 2 \alpha^6 \frac{F''}{F'} \; ,
\end{equation}
and all derivatives of $F$ are evaluated at the background value $\det B = \alpha^6$. 
The speed of sound is smaller than one if $F'' > 0$ and $F' < 0$, in which case the action also has the right overall sign.
All fluids that exist in nature must feature an $F$ with these properties. 
Notice that the transverse modes ($\vec \nabla \cdot \vec \pi = 0$) do not have a gradient energy. As a result, they have a trivial dispersion relation,
$\omega= 0$. They can be thought of as the linear progenitors of vortices \cite{Dubovsky:2005xd}.

In the superluminal case instead, we get the perturbations' action
\begin{equation}
    \mathcal{L} = \alpha^6 F' \Big[ (\partial_1 \pi^1)^2 - (\partial_1 \pi^a)^2 
    -\tilde c^2_s \big(\dot \pi^1 + \partial_a \pi^a\big)^2 \Big] \qquad \qquad \mbox{(superluminal)} \; , \label{nf_x}
\end{equation}
where $a=2,3$ labels the transverse directions, $\tilde c^2_s$ is defined as
\be
\tilde c_s ^2= 1 - 2 \alpha^6 \frac{F''}{F'} \; ,
\ee
and all derivatives of $F$ are evaluated at the new background value $\det B = - \alpha^6$.
As we now show, such a Lagrangian describes a {\it constrained}, {\it unstable} system. 

To see this, consider first decomposing $\pi^a$ into its transverse and longitudinal part w.r.t~to the $x^{2,3}$ derivatives $\partial_a$:
\be
\pi^a = \pi_L^a +  \pi_T^a \; , \qquad \pi_L^a = \partial_a \phi \; , \qquad  \partial_a \pi_T^a = 0 \; ,
\ee
for some function $\phi(x)$
Then, the transverse field $\pi_T^a$ enters the Lagrangian above through the combination $(\partial_1 \pi_T^a)^2$. Its \ equation of motion is thus the constraint
\be
\partial_1^2 \pi_T^a = 0 \; ,
\ee
with independent solutions $f^a_T(t,x^a)$ and $x^1 g^a_T(t,x^a)$, where $f^a_T$ and $g^a_T$ are arbitrary transverse functions. The large space of solutions is a consequence of a gauge invariance, which is itself a consequence -- in this frame and for this peculiar background -- of the original volume preserving diff symmetry. 

Regardless of the solution one chooses, at this order  $\pi^a_T$ is decoupled from the other degrees of freedom, and so, as far as their dynamics is concerned, we can just ignore it. For the reduced $(\pi^1, \pi^a_L)$ system, the above Lagrangian corresponds to the kinetic matrix (up to an overall factor)
\be
K_{IJ} = \left( \begin{array}{cc}
 - \tilde c_s^2 \, \omega^2 + k_\parallel^2 & \tilde c_s^2 \, \omega k_\perp\\
 \tilde c_s^2 \, \omega k_\perp &-k_\parallel^2 - \tilde c_s^2 \, k_\perp^2
 \end{array} \right) 
\ee
where we have expanded the fields in Fourier modes $e^{-i (\omega t + k_\parallel x^1 + k_\perp^a x^a)}$.
The eigenmodes are most easily studied in the long wavelength limit for the $x^1$ direction:
\be
k_\parallel \ll k_\perp, \omega \; .
\ee 
Indeed, in this limit the eigenvalues of $K$ take the form
\be
\lambda_1 = \tilde c_s^2 (\omega^2 + k_\perp^2) + {\cal O}(k_\parallel^2) \; , \qquad 
\lambda_2 = - \frac{\omega^2 - k_\perp^2}{\omega^2 + k_\perp^2}k_\parallel^2 + {\cal O}(k_\parallel^4) \; ,
\ee
which means that the eigenfrequencies of the system are
\be
\omega_1^2 = - k_\perp^2 + {\cal O}(k_\parallel^2) \; , \qquad \omega_2^2 =  k_\perp^2 + {\cal O}(k_\parallel^2) \; .
\ee
Clearly, the eigenmode associated with $\omega_1$ corresponds to an exponential instability, $e^{k_\perp t}$, which is dominated by  UV physics, since it is faster and faster at shorter and shorter wavelengths, signaling a breakdown of the long distance effective theory.

We conclude that, for normal fluids, there cannot be a consistent solution describing the analog of the superluminal vortex that we found in the superfluid case.

\section{Finite temperature superfluids}

We may ask whether heating up a superfluid to some finite temperature has any effect on the conclusions that were drawn in the zero-temperature case. This is an interesting question given  
the usual two-fluid picture for superfluids at finite temperature, where a thermal background of phonons behaving as a normal fluid coexists with a superfluid component. The low-$T$ EFT for such a system was first derived in \cite{Carter:1995if} and the general EFT description was formulated by \cite{Nicolis:2011cs}. As shown in \cite{Kourkoulou:2022doz}, this low-$T$ EFT accurately captures the thermodynamics of the phonons around a general superfluid background $V^\mu$, even when this is spacelike. At the same time, it also allows us to consider any background for the normal fluid component. Having an explicit form for the low-$T$ EFT we are able to check directly, for all different background choices, to what extent the stability properties change when considering a mixture of fluid and superfluid components.

Concretely, we start with the low-$T$ expansion \cite{Kourkoulou:2022doz, Nicolis:2011cs}
\begin{equation}\label{eq:finT}
    \mathcal{L}_0 + \mathcal{L}_1 = P(X) - 3  \bigg[ \frac{b^4 }{ c_s(X) } \Big(1- \big(1-c_s(X)^2 \big) \frac{y^2}{X} \Big)^{2} \bigg]^{1/3},
\end{equation}
where 
\begin{equation}\begin{split}
    c^2_s(X) &\equiv \frac{P'(X)}{P'(X) + 2P''(X) X},\\ y &\equiv \frac{1}{b}\epsilon^{\mu\alpha\beta\gamma}\partial_\mu \psi \partial_\alpha \phi^1 \partial_\beta \phi^2 \partial_\gamma \phi^3 = u^\mu \partial_\mu \psi,\\ b &\equiv \sqrt{B}=\sqrt{\det(\partial_\mu \phi^I \partial^\mu \phi^J)}.
    \end{split}
\end{equation}
 
Here the temperature is measured in the frame determined by the fluid velocity $u^\mu$, and $b$ has the interpretation of entropy density (up to normalization \cite{Kourkoulou:2022doz}), scaling like $b\propto T^3 \ll \mu^3 \ll m^3$. Let us consider $u^\mu = (1,\vec{0} \, )$ to begin with. This corresponds to a stationary and homogeneous fluid background of the form $\phi^I = \alpha (x^I + \pi^I)$, so that $b=\alpha^3$. Considering fluctuations around a generic superfluid background as well, $\psi = V_\mu x^\mu + \pi$, the building blocks of \eqref{eq:finT} are expanded to  
\begin{align}
b^2 &=\alpha^6 (1 +2 \vec\nabla \cdot\vec{\pi}+(\vec\nabla \cdot\vec{\pi})^2 - \dot{\vec{\pi}}^2),\label{expb} \\
X&= -\partial^\mu \psi \partial_\mu \psi = -V^\mu V_\mu - 2 V^\mu \partial_\mu \pi - \partial^\mu \pi \partial_\mu \pi, \label{expx}\\
b y 
    &= b\, u^\mu V_\mu + \alpha^3 V_0 \, \vec\nabla \cdot\vec{\pi} - \alpha^3 \vec{V}\cdot \dot{\vec{\pi}} +b \, u^\mu \partial_\mu \pi \; ,
\label{expy}\end{align}
where we have taken the liberty to integrate by parts some of the quadratic terms, which is allowed since we will be expanding the action up to quadratic order only.

In order to study the system's stability, we plug the above expressions into \eqref{eq:finT} and keep only the leading terms in the low-temperature expansion, up to second order in the fluctuation fields. We will first consider two special cases while keeping a `completely timelike' fluid background, $u^\mu = (1,\vec{0} \, )$: (a) the superfluid background being also completely timelike, $V_\mu \propto (1,\vec{0} \, )$, and (b) the superfluid background being completely spacelike, $V_\mu \propto (0,1,0,0 )$. 


\subsection*{Fluids at relative rest}
Considering first a timelike, stationary superfluid $V_\mu = \mu(1,\vec{0})$, we find
\begin{equation}
    \mathcal{L}_1 = \frac{2\alpha^4}{c_s}\left[ \dot{\vec{\pi}}^2-\frac{c_s^2}{3}(\vec\nabla \cdot\vec{\pi})^2 +\frac{2(c_s^2-1)}{\mu}\dot\pi (\vec{\nabla}\cdot\vec{\pi})-\frac{(c_s^2-1)}{\mu}(\vec\nabla \pi)^2 \right].
\footnote{In complete generality, the coefficients of the structures appearing here are corrected also by terms with a further relative suppresion of $(\mu^2/m^2)$, or higher. We omit these here for a clearer presentation of results, since they don't affect the analysis and conclusions.}
\end{equation}
This correction to the zero temperature Lagrangian includes the regular fluid part, a mixing term, and a new contribution to the superfluid phonon's gradient term -- however the latter two are suppressed and don't affect the stability of the system. To see this more easily we may normalize all the fields canonically, in which case: 
\begin{equation}
\mathcal{L}_0+\mathcal{L}_1 = \frac{1}{2}\left(\dot\pi_c^2 -(c_s^2+ 4(c_s^2-1)\epsilon^2)(\vec\nabla \pi_c)^2 + \dot{\vec{\pi}}_c^2 -\frac{c_s^2}{3} (\vec\nabla \cdot\vec{\pi}_c)^2\ + 4(c_s^2-1)\epsilon\;\dot\pi_c (\vec\nabla \cdot\vec{\pi}_c) \right) , \label{canonical L0+L1}
\end{equation}
where $\epsilon^2\equiv \frac{\lambda \alpha^4}{\mu^2m^2}\ll 1$. The system with $\epsilon = 0$ is strictly subluminal and stable. By continuity, introducing a small $\epsilon$ cannot change that. To see this more explicitly, one can solve for the eigenfrequencies of the system. As before, the transverse components of $\vec \pi_c$ have degenerate dispersion relations, $\omega_T= 0$. On the other hand, the longitudinal component mixes with the superfluid phonon $\pi_c$.
The corresponding eigenfrequencies, for so-called first and second sound, are: 
\begin{equation} 
\omega_1^2 = \left(c_s^2  - 2 \epsilon^2(1-c_s)(3c_s-1)\right)k^2 \qquad
 \omega_2^2 = \left(\frac{c_s^2}{3}-2\epsilon^2(1-c_s)^2\right)k^2,  
 \end{equation}
The propagation speeds then are both subluminal, since so is the original sound speed, $c^2_s \approx 1 - 2(\mu/m)^2 < 1$. We can conclude, then, that the leading effect of heating up the superfluid, in the case where both backgrounds are taken timelike and stationary, is to introduce a normal fluid component with a speed of sound equal to $c_s/\sqrt{3}$. As we just argued, the mixing terms, suppressed by a factor of at least $\epsilon$, do not introduce an instability to the system. We can summarize this as follows:
\begin{equation} 
\mathcal{L}_1 \to \frac{2\alpha^4}{c_s}\left[ \dot{\vec{\pi}}^2-\frac{c_s^2}{3}(\vec\nabla \cdot\vec{\pi})^2\right] .
\end{equation}
To avoid clutter, we will use the same notation in the setups that we consider next, i.e. we will not explicitly write down suppressed terms, unless there are subtleties. Instead, we will report only the leading contributions. 

\newpage
\subsection*{Purely spacelike superfluid}
We now consider instead $V_\mu = \mu(0,\hat {V})$, and in particular choose the background to align with the $x$-direction for convenience. In this case we find:

\begin{equation}
    \mathcal{L}_1 \to \frac{2\alpha^4}{c_s^{\;1/3}}\left[c_s^2 (\dot{\pi}^1)^2 + (\dot{\pi}^2)^2 + (\dot{ \pi}^3)^2- \frac13 (\vec{\nabla}\cdot \vec{\pi})^2\right].
\end{equation}
This is again a contribution from the normal fluid component, but, in this case, anisotropic. There are only compressional modes,  since the only gradient energy term is $(\vec{\nabla}\cdot\vec{\pi})^2$, but the speed of sound depends on the direction. It's worth noting that, with this specific choice of $V_\mu$, the superfluid phonon in $\mathcal{L}_0$ propagates at the speed of light in the $y,z$ directions and has a speed $c^2\approx 1-2(\mu / m)^2$ along $x$. As for the normal fluid fluctuations, we see that both $\pi^2$ and $\pi^3$ propagate with second sound speed equal to $1/ \sqrt{3}$, which is the same fraction of the superfluid speed of sound in those directions as in the previous case. On the other hand, $\pi^1$ has a speed of sound of $1/(\sqrt{3} c_s)$, which in the regime of validity of the EFT  is smaller than one. Hence, also in this case, there are no superluminal fluctuations or other pathologies.

%

\subsection*{Cases where the fluid background is spacelike}

 While the thermal interpretation of this system is physical for timelike normal fluid backgrounds, the form of \eqref{eq:finT} allows us to check what happens in the case where the normal fluid background is taken to be spacelike as well. In this case the question is reversed, i.e.~we want to see whether the instability discussed in the normal fluid section is present here too, or whether the mixing at finite $T$ in $\CL_1 = f(b,y,X)$ somehow stabilizes the system. We have found that to leading order this doesn't happen, and we discuss three simple cases to outline how this conclusion manifests itself: (a) one case where the superfluid background is taken timelike, and two cases where the superfluid background is taken spacelike either (b) in the same or (c) in a perpendicular direction with respect to the (spacelike) normal fluid background.

In analogy with the examples above, we perform variable expansions for $X, b ,y$ as in \eqref{expb}-\eqref{expy} with the field configurations that suit each case, and plug these modified expressions into the low-$T$ Lagrangian \eqref{eq:finT}. 


\begin{enumerate}[\textbf{(\alph*)}]
\item Here we choose a stationary, timelike superfluid background $V_\mu$ like before, and a spacelike background in the $x$-direction for the normal fluid. The corrections to the Lagrangian are:
\begin{equation}\label{eq:tx}
    \mathcal{L}_1 \to \frac{2\alpha^4}{c_s^{\;1/3}}\left[c^2_s (\partial_1 \pi^1)^2- (\partial_1 \pi^2)^2-(\partial_1 \pi^3)^2-\frac13(\dot{\pi}^1 + \partial_2\pi^2+\partial_3 \pi^3)^2\right].
\end{equation}
This takes the same form we encountered in \eqref{nf_x}; a similar analysis would reveal the same type of exponential instability. There is a mixing term that appears to subleading order of the form $\partial_1\pi\,\partial_1\pi^1$, further suppressed by a factor of $\alpha^2/(\mu m)$, but it does not remove the instability.
\item Taking both backgrounds to be spacelike and parallel (in the $x$-direction for simplicity), we find:
\begin{equation}
     \mathcal{L}_1 \to \frac{2\alpha^4}{c_s}\left[(\partial_1 \pi^1)^2-(\partial_1 \pi^2)^2- (\partial_1\pi^3)^2-\frac{c_s^2}{3}(\dot{\pi}^1 + \partial_2 \pi^2 + \partial_3 \pi^3)^2\right].
\end{equation}
Up to multiplicative factors, this correction is of the same form we found in the case above, and so the same type of instability carries over.
\item Finally, we take both the superfluid and normal fluid backgrounds to be spacelike, pointing in orthogonal directions $x$ and $y$. We find
\begin{equation}
    \mathcal{L}_1 \to \frac{2\alpha^4}{c_s^{\;1/3}}\left[(\partial_2 \pi^2)^2-c_s^2(\partial_2 \pi^1)^2-(\partial_2 \pi^3)^2- \frac13(\dot{\pi}^2 +\partial_1 \pi^1+\partial_3 \pi^3)^2\right].
\end{equation}
 Perhaps unsurprisingly by now, we observe the same structure, with the expected minor twist of the fluctuation along the direction of the fluid background being the dynamical one (in this case $\pi^2$, as opposed to $\pi^1$ in the previous examples). The exponential instability characteristic of spacelike backgrounds for a normal fluid is present here too.
\end{enumerate}

For clarity, we have only presented the simplest cases that capture all the possible distinct arrangements one could consider, but one can generalize to arbitrary backgrounds (as opposed to picking a direction, like we did in these examples). 
We have checked this explicitly in the physically relevant scenario, where the normal fluid background is chosen timelike, but moving. The expressions look too messy to claim that their presentation in this paper would be illuminating to the reader, but they lead to the same conclusion: that the finite temperature superfluid retains the peculiar property of being stable and having only subluminal excitations, even for spacelike background configurations, as long as the normal fluid component is timelike.


\newpage
\section{Velocity fields in thermodynamics}

At the level of the effective action, the two four-velocities -- describing the motion of the normal fluid and superfluid -- are merely parameters that appear in the constitutive relations for the Noether currents, namely the stress-energy tensor and particle-number currents. They are vectors that pick out a preferred local rest frame of the medium or some sub-component of the medium. From this perspective, the four-velocity of the fluid and superfluid appear to be on the same footing. So why should it be that in the examples we considered, a superluminal superfluid velocity is okay while a superluminal normal fluid velocity is unstable and hence unphysical? The answer lies in the fact that our effective action approach obscures the fundamental physical differences between these two notions of four-velocity. 

\subsection*{Normal fluids}

Begin by considering normal fluids. By their nature, they are inherently thermal media; no zero-temperature ordinary fluids exists. 
Their equilibrium state is characterized entirely by the thermal density matrix
\begin{equation}\label{density 0}
     \rho_0^{\rm fluid} = \frac{e^{-\beta H}}{Z(\beta)}  \; ,\qquad Z(\beta) \equiv \tr e^{-\beta H},
\end{equation}
where $\beta$ is the equilibrium inverse temperature and $H$ is the Hamiltonian,  $H=P^0=-P_0$, where $P_\mu$ are the generators of spacetime translations. (For simplicity, we are assuming that the only symmetries present are Poincar\'e transformations, i.e.~there are no chemical potentials.)   

We would like a covariant expression for the thermal density matrix. To this end, we must introduce a constant time-like four-vector $\beta^\mu$, which can be decomposed into norm and direction by $\beta^\mu = \beta u^\mu$, where $u^\mu$ is a unit vector. The density matrix then takes the form 
\begin{equation}
\rho^\text{fluid}= \frac{e^{\beta^\mu  P_\mu}}{Z(\beta)}  \; ,\qquad Z(\beta) \equiv \tr e^{\beta^\mu P_\mu}. \end{equation} 
To recover the original expression~\eqref{density 0}, all we need do is choose $u^\mu$ to point along the time direction, namely $u^\mu  = (1, \vec 0 \,)$. To work in any other frame, we can suitably boost $u^\mu$. The interpretation of $u^\mu$ is now clear: it specifies the zero-momentum frame. As a result, we identify it as the equilibrium four-velocity of the fluid. 

Assuming that the underlying physics is stable -- that is that $H$ is bounded from below -- and that we appropriately regularize our system by placing it in a finite volume, the density matrix is well-defined so long as $u^\mu$ is subluminal. In this case, we may always boost to a frame in which $u^\mu = (1, \vec 0 \,)$, meaning that $u^\mu P_\mu = - H$. Explicitly, the expression for the partition function $Z$ is therefore
\begin{equation}
Z(\beta) = \int dE  \, \varrho_0(E) \, e^{-\beta E},
\end{equation}
where $\varrho_0$ represents the density of states with energy $E$. For any stable system, the Hamiltonian is bounded below by some energy $E_\text{min}$. Thus the density of states satisfies $\varrho_0(E)=0$ for any $E< E_\text{min}$, so no divergence in the definition of $Z$ can arise because of negative-energy states. The only way for $Z$ to be ill-defined is if $\varrho_0(E)$ grows sufficiently fast for large $E$ (or if $\varrho_0$ diverges badly enough for some value of $E$). But so long as $\varrho_0(E)$ is finite and grows slower than $e^{\beta E}$, which is a reasonable assumption for most situations, then $Z$ is  well-defined. As a result, the density matrix is well defined. 

Now suppose that $u^\mu$ is spacelike. Then, we can always boost to a frame in which $u^\mu$ has no temporal component. Without loss of generality suppose that $u^\mu = (0,1,0,0)$. Then the density matrix takes the form
\begin{equation}\rho_x^\text{fluid} = \frac{e^{\beta P_x}}{Z_x(\beta)}  \;, \qquad Z_x(\beta) \equiv \tr e^{\beta P_x}.  \end{equation}
In terms of the density of states of definite $x$-momentum $p_x$ given by $\varrho_x(p_x)$, the explicit expression for the partition function is
\begin{equation} Z_x (\beta)= \int dp \,  \varrho_x(p)  \, e^{\beta p}.  \end{equation}
Our situation now is fundamentally different from the case in which $u^\mu$ was subluminal. Unlike the Hamiltonian, the momentum operator $P_x$ is not bounded from above or below. In fact, even in theories without parity, by rotational invariance the spectrum of $P_x$ must be symmetric around zero. As a result, the only way for $Z_x$ to be finite is if $\varrho_x(p)$ decays faster than $e^{-\beta p}$ for large $p$. This behavior for the density of state is highly unrealistic; indeed it is impossible for relativistic systems. As a result, $Z_x$ is divergent, meaning that $\rho_x$ is non-normalizable and hence ill-defined. Thus $\rho^\text{fluid}$ does not exist when the fluid four-velocity is superluminal, so it should be no surprise that superluminal fluid four-velocities give nonsensical results in the effective field theory. 

\subsection*{Superfluids}

Now consider a zero-temperature superfluid; the description in terms of a thermal density matrix is still possible, as a $\beta \to \infty$ limit, but not necessary. The equilibrium state of the superfluid is represented by a pure quantum state 
that spontaneously breaks various symmetries, typically Poincar\'e symmetry and a $U(1)$ charge corresponding to particle-number conservation. Denote spacetime translation generators by $P_\mu$, Lorentz symmetry generators by $J_{\mu\nu}$ and the $U(1)$ generator by $Q$. The ground state of the system $\ket \Omega$ spontaneously breaks $P_0$, $Q$, and $J_{0i}$, while leaving $J_{ij}$, $P_i$, and the diagonal combination $\bar P_0 \equiv P_0 + \mu Q$ unbroken, as clear from the background field $\psi(x)=\mu t$. We interpret $\mu$ as the equilibrium chemical potential, and $\bar H = \bar P^0 = H- \mu Q$ as the Hamiltonian for the superfluid.  In particular, the ground state of a superfluid is the lowest lying eigenstate of $\bar H$. A superfluid exists for some $\mu$ only if the corresponding $\bar H$ is bounded from below.

Now consider a different state for our superfluid, one in which the background field is not purely timelike, like for instance the one in \eqref{background Vmu}. In this case, the unbroken linear combinations of $P_\mu$ and $Q$ are
\be \label{P bar}
\bar P_\mu =  P_\mu + V_\mu Q \; .
\ee 
In particular, the ground state of the superfluid will be the lowest lying eigenstate of 
\be \label{H bar}
\bar H = \bar P^ 0 = H - V_0 Q 
\ee
that is also an eigenstate of $\bar P_i = P_i + V_i Q$. If $V_\mu$ is timelike, this will simply be the boosted version of the superfluid's ground state we had above, for $V_\mu = (\mu, \vec 0 \,)$. If on the other hand $V_\mu$ is spacelike, which would suggest a superluminal superfluid velocity, this will be another state, but we cannot see any obvious pathologies associated with it. After all, \eqref{H bar} is bounded from below, at least in a range of values for $V_0$, because of the assumptions above.

%
Whenever $V_\mu$ is spacelike we may choose a frame in which it has no component along the time direction. Without loss of generality, suppose that $V_\mu  = (0,V_x,0,0)$. Then the unbroken generators are
\begin{equation}\label{SSB pattern 0} 
P_\text{a} \; , \qquad P_x+ V_x Q \; , \qquad J_\text{ab} \;   ,  \end{equation}
where $\text{a,b} = 0,2,3$, while the broken generators are  $P_x$, $Q$, and all Lorentz transformations that act non-trivially on $V_\mu \propto \delta^x_\mu$. 
This symmetry-breaking pattern is most unlike those of normal states of matter \cite{Nicolis:2015sra}. In particular, we usually expect a medium to spontaneously break all boost symmetries, while here boosts about the $y$- and $z$-directions remain unbroken. Although unusual, there is nothing intrinsically pathological about this SSB pattern. Whether or not it is possible depends on the particular model in question. As a result, it should not come as a surprise that there exist certain superfluid theories that admit apparently superluminal four-velocities, in the sense that the associated $V_\mu$ is spacelike. 

We should note, however, that calling such a state of matter a `superfluid' is not entirely correct. Although we began with the action for a superfluid, as soon as $V_\mu $  becomes spacelike, the spontaneous symmetry breaking pattern becomes fundamentally different. It is therefore best to conceptualize the superluminal core of our superfluid vortex as no longer existing in superfluid phase. Instead it exists in the heretofore unnamed phase of matter characterized by the above SSB pattern. We will see that at finite temperature, all boosts are once again broken and this phase of matter has a name. 

\subsection*{Finite temperature superfluids}
Ignoring the usual subtleties about how to properly characterize SSB for a thermal density matrix \cite{forster1975hydrodynamic}, we write the density matrix
\be
     \rho^\text{super} =  \frac{e^{\beta^\mu \bar P_\mu}}{Z(\beta, V)}\; ,\qquad 
     Z(\beta, V) = \tr e^{\beta^\mu \bar P_\mu} ,
\ee
with the same notation as above for $\bar P_\mu$ (see eq.~\eqref{P bar}). Such a density matrix describes the thermodynamics of a fluid system with a conserved charge. For certain values of the control parameters $\beta^\mu$ and $V_\mu$, such a charge will be spontaneously broken and in that case we  end up with a finite-temperature superfluid.
%
We now want to understand under what conditions on $\beta^\mu = \beta u^\mu$ and $V_\mu$ we can expect the density matrix to be normalizable.
 
First, suppose that $u^\mu$ is timelike and choose coordinate such that $u^\mu = (1, \vec 0 \,)$. We get 
\be
     \rho^\text{super}_0 = \frac{e^{-\beta( P^0 - V_0 Q)}}{Z(\beta, V)} \; , \qquad Z(\beta, V) = \tr e^{-\beta( P^0 - V_0 Q)} \; .
\ee
Such a density matrix is well-defined assuming that $P^0- V_0 Q$ is bounded from below, which is a prerequisite assumption for the superfluid phase to exist. This is essentially the same conclusion that we had for the zero-temperature superfluid.
Thus no matter the choice of $V^\mu$, as long as $u^\mu$ is timelike and the desired SSB pattern can be realized at temperature $1/\beta$, the density matrix is well-behaved. 

Next, suppose that $u^\mu$ is spacelike and choose coordinates such that $u^\mu  = (0,1,0,0)$. The density matrix then takes the form
\be
     \rho^\text{super}_x = \frac{e^{\beta( P^x + V_x Q)}}{Z(\beta, V)},\qquad Z(\beta, V) =  \tr e^{\beta( P^x + V_x Q)} \; .
\ee
As was the case for the normal fluid, $P^x$ is totally unbounded -- the same will be true for $P^x+V_x Q$ in a sector with fixed finite charge. To see how this is so, consider a state of fixed finite charge and non-zero 4-momentum. By Lorentz symmetry, this state can be boosted along the $x$-direction to have arbitrary momentum $p_x$. As the charge is unaffected by boosts, we see that $P^x + V_x Q$ is totally unbounded. 
Thus, when $u^\mu$ is spacelike, the density matrix can never be normalized. 

Notice that the claim we are making here is stronger than what we showed by explicit calculations in previous sections. We previously considered special cases for the fluid and superfluid velocity within a particular model. In all such cases, a superluminal normal fluid velocity spelled catastrophe. Here, we have demonstrated that for {\it any}  fluid or superfluid state in a relativistic theory at finite temperature, a superluminal normal fluid velocity leads to a pathologically non-normalizable state and it is hence unphysical. 

\newpage
To summarize: the normal fluid velocity must remain subluminal or else the density matrix fails to be normalizable, while a superfluid four-velocity has no such restrictions. When the superfluid four-velocity becomes superluminal, however, the resulting SSB pattern no longer describes a superfluid phase. For the simplified case in which $u^\mu $ is along time and $V_\mu $ is along $x$, the broken generators are $P_x$, $Q$, $J_{0i}$, and $J_{xi}$, while the unbroken generators are temporal translations, spatial translations orthogonal to $x$, and the diagonal combination $P_x+V_x Q$. This SSB pattern gives rise to a state of matter with a well-known name: smectic liquid crystal in phase-A \cite{Landry:2020ire,PhysRevA.6.2401}. This state of matter looks like a solid along one direction, that is $P_x$ is broken, but looks like a liquid along the remaining two directions as $P_y$ and $P_z$ are unbroken. Unlike ordinary smectic liquid crystals, however, the superluminal superfluid exhibits two longitudinal sound modes. This appearance of second sound at low temperatures is a general phenomenon also in solids, which arises, roughly speaking, due to the exponential suppression of Umklapp scattering \cite{pitaevskii}.

\section{Relation to giant vortices}
Ref.~\cite{Cuomo:2022kio} studied a system that has some relationship with ours. That paper considers a conformal superfluid EFT described, in 2+1 dimensions, by the action
\begin{equation}
    P(X) = X^{\frac32} \; .
\end{equation}
On the sphere, for a strip around the equator, they  consider what they call a {\it giant vortex} solution 
\begin{equation}
    \bar{\psi} = \mu t + \ell \varphi \; ,
\end{equation}
which we recognize as our superluminal vortex, now allowing for more general angular momentum, parametrized by $\ell$. 
They find the presence of chiral modes moving at the speed of light, and argue that these modes are what allows the giant vortex to have rapid rotation. The approximations they make to find these modes are similar to zooming in near the $X=0$ region in our setup. So, let's try to do that in our case. Focusing on fluctuations in a small strip near $r_* = \mu^{-1}$, and on modes which only depend on $\varphi$ and $t$, we get
\begin{equation}\label{eq:llback}
    \lambda \CL_\pi = \mu^2 (\partial_- \pi)^2 + \frac{m^2}{2}\partial_+\pi\partial_- \pi,
\end{equation}
where $\partial_{\pm} = \partial_t \pm \mu \partial_\varphi$. We thus find one mode, in the $-$ direction, moving at speed of light, whereas the $+$ mode is almost, but not quite, lightlike, since it moves at $c = \frac{m^2-2\mu^2}{m^2+2\mu^2} \simeq 1 - 4 \frac{\mu^2}{m^2}$. These are essentially the chiral modes of \cite{Cuomo:2022kio}. The main difference is that for us the relative coefficient $m^2/\mu^2$ in \eqref{eq:llback} is large, whereas in \cite{Cuomo:2022kio} it is instead very small (there is no analog of $m$ for a conformal superfluid).

\newpage
\section{Discussion}

Our results cast doubts on the interpretation of a superfluid velocity field as the actual velocity field of some form of matter. As we briefly discussed in the Introduction, to understand the issue better one should probably be  more precise in defining what it means -- in a quantum theory and in particular in relativistic QFT -- that something physical is moving at a given speed.

Even without doing so, we can make some sense of our results. Recall that we trust our superluminal vortex solution within the superfluid EFT only for cases in which we have SSB already at vanishing chemical potential, so that we can have a superfluid phase at arbitrarily low values of the chemical potential. This makes our superfluid somewhat peculiar compared to, say, helium-4: there, the absolute ground state of the theory is the standard Poincar\'e-invariant, $U(1)$-invariant vacuum. Then we have excited states that are approximately described by well-separated particles (i.e., helium-4 atoms), and only at nonzero density, for chemical potentials that exceed the mass of a single particle, do we break the $U(1)$ symmetry and form a superfluid. Given this physical picture, we can think of superfluid helium-4 as being made up of helium atoms, and we can characterize its  velocity field as some sort of local average of its constituents' velocities.

On the other hand, in our case we have no such interpretation available: our $U(1)$ symmetry is always spontaneously broken, and so we don't have the analog of the helium atoms, each carrying one unit of charge. In a sense, our superfluid is not made up of particles. It is a more field-theoretical object, which at finite chemical potential has the same symmetries and symmetry-breaking pattern as an `ordinary' superfluid, but with a strikingly different starting point.

It might well be that superfluids of this sort, whose SSB survives all the way to vanishing chemical potential, should be interpreted as {\it generalized superfluids}, with only a formal, symmetry-based connection to ordinary superfluids made up of particles. It might be that the resolution of our superluminal puzzle is that for ordinary superfluids like helium-4 the superfluid velocity field is the physical velocity of `something', and is thus constrained to be subluminal, whereas for our generalized superfluids there is `nothing' moving at that speed. Without a better characterization of
`something' and `nothing,' it is difficult to test this idea, but the distinction between generalized and ordinary superfluids might turn out to be physically relevant.
 
\subsection*{Acknowledgements}
We would like to thank Matteo Baggioli for valuable comments, as well as Gabriel Cuomo, Angelo Esposito, and the other organizers and participants of the 2022 Pollica Summer Workshop on Effective Field Theory for stimulating discussions. IK, AN and KP were supported in part by the US DOE (award number DE-SC011941) and by the Simons Foundation (award number 658906). ML was supported in part by the ALFA foundation. 

\newpage
\bibliographystyle{JHEP.bst}
\bibliography{superl}

\end{document}